\documentclass[preprint]{aastex}

\usepackage{natbib}
\usepackage{graphicx}

\slugcomment{To appear in Astrophysical Journal Letters}

\shorttitle{Extreme host galaxy growth}
\shortauthors{Barthel et al.}

\begin{document}

\title{Extreme host galaxy growth in powerful \\ early-epoch radio galaxies}

\author{Peter Barthel}
\affil{Kapteyn Astronomical Institute,
       University of Groningen, The Netherlands}
\email{pdb@astro.rug.nl}

\author{Martin Haas}
\affil{Astronomisches Institut, Ruhr Universit\"at, Bochum, Germany}

\author{Christian Leipski}
\affil{Max-Planck-Institut f\"ur Astronomie, Heidelberg, Germany}

\and

\author{Belinda Wilkes}
\affil{Harvard-Smithsonian Center for Astrophysics, Cambridge, Massachusetts, USA}

\begin{abstract}

During the first half of the universe's age, a heyday of star-formation
must have occurred because many massive galaxies are in place after that
epoch in cosmic history.  Our observations with the revolutionary
Herschel Space Observatory\footnote{Herschel is an ESA space observatory
with science instruments provided by European-led Principal Investigator
consortia and with important participation from NASA} reveal vigorous
optically obscured star-formation in the ultra-massive hosts of many 
powerful high-redshift 3C quasars and radio galaxies. This symbiotic 
occurrence of star-formation and black hole driven activity is in marked 
contrast to recent results dealing with Herschel observations of X-ray 
selected active galaxies. Three archetypal radio galaxies, at redshifts
1.132, 1.575, and 2.474 are presented here, with inferred star-formation
rates of hundreds of solar masses per year.  A series of spectacular
coeval AGN/starburst events may have formed these ultra-massive galaxies
and their massive central black holes during their relatively short
lifetimes. 

\end{abstract}

\keywords{Galaxies: formation --- Galaxies: starburst ---
          Galaxies: high-redshift --- Infrared: galaxies}

\section{Introduction}

The most massive galaxies known have stellar masses $\sim 5 \times
10^{11}$ M$_{\odot}$, and the fact that they already exist at early
epochs \citep{fontana06} implies that they must have formed rapidly
\citep{daddi05}, during just a few Giga-years.  They are expected to
host supermassive black holes \citep{haring04} which have periods of
active accretion \citep{breuck10}.  A symbiotic occurrence of modest
star-formation and accretion activity has been inferred in the high mass
host galaxies of intermediate redshift X-ray selected AGN
\citep{lutz10}.  Additional interest for details of the symbiosis in
radio-loud AGN is moreover generated by the proposed negative AGN
feedback \citep[e.g.,][]{bower06}.  Here we quantitatively address
star-formation and black hole build-up in the most massive galaxies
hosting radio-loud AGN, during their time of formation. 

Powerful high-redshift radio galaxies are ideal tracers of the extreme
form of starburst-AGN symbiosis for several reasons.  Firstly, their
central black holes are being fed at a high rate, hence are rapidly
growing.  Secondly, their large extended radio sources allow an
unambiguous quantification of the accretion power, from the huge radio
luminosities; moreover, their radio morphological properties permit an
estimate of the duration of the AGN and possibly coeval star-formation
episodes.  Thirdly, the presence of large reservoirs of molecular gas,
the necessary fuel for star-formation, has been inferred \citep{solomon05}
from (sub)-millimetre spectroscopy in some radio galaxies
(but not in all).  Quantification of the ongoing star-formation is
provided through far-infrared (FIR) observations, since dust absorbs the
radiation of the young stars and re-emits it at FIR wavelengths.  From
rest-frame submillimetre photometry \citep{benford99, reuland04} and
mid-infrared spectral energy distributions, SEDs \citep{seymour08}
it was suspected that some powerful high redshift radio galaxies might
be experiencing phases of copious (dust obscured) star-formation.  The
implications from the SED measurements, however, were uncertain: the
amount of cool dust reradiating the emission of the newly formed stars
was not well constrained, as the rest-frame far-infrared regions and the
Rayleigh-Jeans tails lack data. 

The Herschel Space Observatory \citep{pilbratt10} provides full
measurement of IR-submm SEDs at unprecedented sensitivity.  This Letter
presents breakthrough results on a sample of seventy radio-loud,
high-redshift AGN, namely quasars and radio galaxies from the 3C and 4C
catalogues.  With P$_{\rm 1.4\,GHz} \sim 10^{27.5}$ W Hz$^{-1}$, these
AGN contain the most powerful accreting black holes.  Inspection of our
sample measurements shows FIR detection fractions of roughly two in
three objects, at both 160\micron~and 250\micron~(from the Herschel
PACS and SPIRE instruments, respectively).  Stressing that the
non-detected objects are probably just below the sensitivity limit, we
focus here on three archetypal objects, reporting on their spectacular
on-going star-formation. 

3C\,368 \citep{best97, best98a, best98b, chambers90, djorgov87}, 
3C\,68.2 \citep{best97, best98a, chambers90, djorgov88} 
and 3C\,257 \citep{breugel98}, at redshifts $z$ = 1.132, 1.575, and
2.474 respectively, are archetypal protogalaxies with well-documented
properties.  3C\,257 is the highest redshift object in the complete 3CR
sample \citep{spin85}. The first studies of 3C\,68.2 and 3C\,368 in the 
mid and late 1980s, and their identification as protogalaxies marked the 
birth of the K--$z$-diagram \citep{lilly84} and the beginning of extensive
observations of their massive hosts in a cosmological context \citep{spin86}.
The presence of a luminous QSO -- a Type-1 AGN, obscured from direct view 
behind a toroidal circumnuclear dust configuration -- was proven in the 
case of 3C\,257 from X-ray data \citep{derry03}; similarly the two other 
objects reveal obscured QSOs, in our as yet unpublished Chandra data. 

\section{Observations and results}

Photometric observations of 3C\,368, 3C\,68.2 and 3C\,257 were carried
out as part of a Herschel Guaranteed Time program -- see Table~1.  The
PACS scan-map mode was employed with cross scans in the blue (70\micron)
and the red (160\micron) bands, yielding angular resolution of
respectively 5\arcsec~and 11\arcsec.  Data reduction was performed
within HIPE \citep{ott10} following standard procedures for deep field
observations, including source masking and high-pass filtering.  Both
scan directions were processed individually and then mosaiced to yield
the final map.  Aperture photometry (including appropriate aperture
corrections) of the sources at the known radio core positions was
carried out to measure source flux densities.  Photometric uncertainties
were determined by measuring the flux in several apertures on empty
parts of the background.  The SPIRE small map mode was employed in the
250\,$\mu$m (18\arcsec~resolution), 350\,$\mu$m (25\arcsec) and
500\,$\mu$m (36\arcsec) bands.  Data processing followed standard
procedures for small map data within HIPE.  The photometry was performed
utilizing a SPIRE source extractor also implemented in HIPE
\citep{sav07}.  The uncertainties of the SPIRE photometry are dominated
by confusion noise.  Combining the Herschel data with our existing
\citep{haas08} Spitzer mid-infrared photometry, we were able to trace
our science targets as well as other objects in the fields over two
decades in wavelength.  With this approach we ensure proper
identifications while avoiding source confusion.  Supplementing our
Herschel and Spitzer data with published photometry we obtain the
complete optical-IR-submm SEDs.  To extract physical parameters, we
fitted the data with three components representing typical constituents
of a radio galaxy SED.  In practice, we started with the AGN-powered
warm dust emission by cycling through a library of torus models covering
a wide range in the relevant parameter space \citep{honig10}.  For each
torus model (using edge-on inclinations of 60\degr, 75\degr, or 90\degr)
we added a black body in the optical-NIR for emission from the host
galaxy stars\footnote{3C\,257 required the addition of an extra black
body of $\sim$1300\,K.  Such a component, which is generally identified
with hot dust close to the sublimation temperature, is often observed in
Type-1 AGN but has not been seen in Type-2s so far.  Inclusion or
exclusion of this hot dust component does not affect the forthcoming
conclusions.} as well as a modified black body of variable temperature
(grey-body model with $\beta$ fixed at 1.6) in the FIR/submm to account
for possible excess emission due to star-formation.  Figure~1 shows for
each source the combination of one of the torus models, scaling for all
components, and FIR dust temperature which overall minimizes the
chi-square, i.e., the best fitting model.  Errors on the derived
physical parameters were determined from the distribution of these
parameter values for the range of model combinations consistent with the
data.  Since radio galaxy jets have large inclinations, radio core
emission is relativistically de-boosted rather than boosted as in
quasars.  Core radio data \citep{best97, best98b, breugel98} indeed show
that any non-thermal contribution to the submm emission is negligible. 
While the SPIRE beam also encloses the radio source hot spots, their
contribution is negligible since even the total integrated radio SED
underpredicts the FIR/submm by a large margin.  As Figure~1 shows,
substantial cool dust emission is called for by the SED grey bodies,
with temperatures of 53\,K, 36\,K and 37\,K, for 3C\,368, 3C\,68.2 and
3C\,257 respectively.  Using standard cosmology (H$_0$ = 70 km s$^{-1}$
Mpc$^{-1}$, $\Omega_{\Lambda} = 0.73$, $\Omega_m = 0.27$) we compute
source intrinsic properties, which are listed in Table~2.  The listed
L$_{\rm AGN}$-values originate from the torus modeling: they represent
the (accretion) luminosities required to power the fitted torus
emission.  The L$_{\rm SF}$-values were determined by integrating the
fitted FIR grey bodies between 8\micron\ and 1000\micron. 

\section{Discussion}

In analogy with other studies \citep{benford99, hatzim10} we attribute 
the torus heating, i.e., the warm dust component, to the obscured AGN: 
their luminosities are $\sim 10^{46}$ erg sec$^{-1}$, confirming the 
luminous active accretion in these radio-loud Type-2 AGN, which is 
consistent with the unification scenario \citep{barthel89, derry03,
haas04, ogle06}. Attributing the cool dust emission to star-formation, 
we determine luminosities which are comparable to the AGN luminosities
(Table~2), all in excess of $10^{12}$ L$_{\odot}$.  Using standard
conversions \citep{robk98}, the star-formation rate (SFR) values are 610, 
390 and 770 M$_{\odot}$ yr$^{-1}$, for 3C\,368, 3C\,68.2 and 3C\,257,
respectively.  Hence, all three high-$z$ radio galaxies are prodigious
star-formers\footnote{For comparison, we performed the same decomposition
for the host of the powerful low-redshift radio galaxy Cygnus~A (3C\,405), 
using published \citep{haas04} SED data, finding considerably lower values:
L$_{\rm AGN} = 8.2 \times 10^{44}$ erg sec$^{-1}$ and
L$_{\rm SF} = 1.5 \times 10^{11}$ L$_{\odot}$ 
(SFR 26 M$_{\odot}$ yr$^{-1}$).}.
Four points are noteworthy: (1) when comparing the SFRs with values
inferred from uv/optical data \citep{chambers90} we conclude that most 
of the star-formation in these three objects is strongly obscured in
that band, 
(2) as judged from their radio sizes (this point will also return later), 
these are mature objects rather than young AGN in a transition phase 
from dust obscured infrared galaxies to AGN, (3) 3C\,68.2 and 3C\,257 
have cool dust temperatures in the range of those measured for distant
sub-millimetre galaxies \citep{magnelli12}, and (4) as judged from the
detection statistics ($\sim 70\%$), many (if not all) of 
our sample objects must have optically obscured star-formation of 
comparable strength, i.e., SFRs of hundreds of M$_{\odot}$ yr$^{-1}$. 
The high star-formation luminosities for these high-$z$ radio galaxies 
are consistent with results from deep field Herschel surveys targeting 
a mix of radio-quiet AGN types over a large redshift range out to 
$z \sim 3$ \citep{hatzim10, elbaz10, mullaney12a} and with results 
from $z > 4$ QSO studies \citep{leipski10}, but our detection fraction
is higher.

Our observations indicate that many distant powerful radio galaxies
display star-formation at a level comparable to ultra-luminous infrared
galaxies (ULIRGs) which is coeval to their black hole activity.  That
conclusion together with the above mentioned results of Herschel studies
dealing with powerful radio-quiet AGN are in marked contrast with the
conclusion \citep{page12} for X-ray selected $1 < z < 3$ (radio-quiet)
AGN in the CDF-N, namely that ultra-luminous starburst activity is not
seen in the most powerful AGN.  While detailed comparison of these
conflicting results is beyond the scope of this Letter we point out two
possible explanations.  Firstly, the hosts of the CDF-N AGN are of lower
stellar mass than the radio source hosts.  We stress that distant 3C and
4C objects (of which only of order $10^2$ are known) represent the
highest peaks in the galaxy mass distribution.  In fact, our
observations are in broad agreement with the extrapolated cosmologically
evolving specific SFR, sSFR, in the hosts of luminous AGN (and
star-forming galaxies) recently established by \citet{mullaney12a}.  A
-- puzzling -- inverse correlation between AGN X-ray strength and host
galaxy mass would be required to explain the \citet{page12} results
within the just mentioned sSFR behaviour in combination with our
findings for radio-loud AGN.  Secondly, radio-selected AGN -- selected
using radio lobe luminosity which is an isotropic property -- represent
a cleaner, more complete AGN subsample than do X-ray-selected AGN: the
latter may lack the most obscured sources.  The X-ray luminosities
reported by \citet{page12} are uncorrected for obscuration, who estimate
the effect is small.  X-ray observations for the 3C sources reported
here indicate observed luminosities, assuming a standard power-law
spectrum with $\Gamma = 1.9$, of about $3 \times 10^{43}$ erg s$^{-1}$,
$4 \times 10^{43}$ erg s$^{-1}$ (Wilkes et al., 2012, in prep.) and $1.5
\times 10^{44}$ erg s$^{-1}$ \citep{derry03} for 3C\,68.2, 3C\,368 and
3C\,257 respectively, comparable with sources in the \citet{page12}
sample.  It is well known that low signal-to-noise (S/N) X-ray data
provide poor estimates of obscuration and so result in luminosities
uncertain by 1--2 dex \citep{cappi06, wilkes05}.  Indeed, for all three
3C sources reported here (radio galaxies), the X-ray data indicate
significant obscuration with estimated N$_{\rm H} \sim 10^{23-24}$,
yielding estimated intrinsic, hard-band X-ray luminosities of $\sim
10^{45}$ erg s$^{-1}$ for 3C\,68.2 and 3C\,368 (Wilkes et al., 2012, in
prep.) and $9 \times 10^{44}$ erg s$^{-1}$ for 3C\,257 (2--10 keV,
\citet{derry03}).  The low S/N of the X-ray data combined with the
likely presence of additional, unobscured, radio-jet-linked emission
(responsible for the typically $3 \times$ brighter X-ray luminosities in
radio-loud quasars, \citet{zamor81}) make it likely that the N$_{\rm H}$
and intrinsic X-ray luminosity for these three 3C sources remain
underestimated.  Although X-ray emission in radio-quiet AGN is less
complex and may be somewhat lower luminosity, the stronger bias against
highly-obscured sources and the difficulty in detecting absorption in
the fainter X-ray sources, which are those most likely to be obscured
\citep{kim07}, is very likely to result in significantly underestimated
X-ray luminosities in the Page et al.  X-ray-selected sample.  The fact
that a higher incidence of X-ray absorption in those sources with
250$\mu$m detections is also reported strengthens this possibility and
so questions their main conclusion: that 250 $\mu$m emission is
undetected, indicating that SF is quenched, in the highest luminosity,
L$_X \gtrsim 10^{44}$ erg s$^{-1}$, active galaxies in their sample. 
Careful examination of the true nature of the high-$z$ X-ray AGN is
required to back-up the proposed negative feedback; such feedback is
definitely not observed in the radio-loud(est) AGN.  While the duration
of the vigorous star-formation phase could be longer than the radio-loud
activity episode, with the star-formation starting earlier than and
possibly eventually being quenched by the AGN (analysis of our full
sample of young, mature and old radio sources will address that issue),
based on our observations there is a substantial symbiotic period. 

In contrast to other AGN, radio galaxies -- with their extended radio
structure at high inclination -- offer the unique possibility of
estimating the age of the activity episode from the linear size of the
structure, adopting a typical \citep{best95} source expansion speed of ten 
to twenty per cent of the speed of light.  The average duration of such an
episode follows from the maximum size of radio galaxies at the relevant
cosmic epoch.  Then 3C\,68.2, measuring $\sim$190kpc, has a large
dimension implying its AGN phase has an age of several million years,
while the sizes of 3C\,368 and 3C\,257 indicate that they are
somewhat younger.  Focussing on 3C\,68.2, and adopting an AGN age of 
5~Myr, a symbiotic star-formation phase at the inferred SFR of 
$\sim$500 M$_{\odot}$ yr$^{-1}$, assuming it to be constant, would 
yield $\sim2.5 \times 10^9$ M$_{\odot}$. The actual figure could 
be larger if the star-formation commenced before the AGN phase.  The 
host mass of 3C\,68.2, $5 \times 10^{11}$ M$_{\odot}$ \citep{best98a}, 
can be formed through $\sim$200 such gas accretion phases, for which 
there is sufficient time given the age of the universe at that redshift 
of 4~Gyr.  The $\sim$200 associated AGN phases will last a total time of 
$\sim10^9$ yr during which a massive central black hole of 
$\sim1 \times 10^9$ M$_{\odot}$ must accrete \citep{haring04}.  
The implied average black hole fuel consumption of 
1 M$_{\odot}$ yr$^{-1}$ yields  an energy output 
$\sim\eta$.dm/dt.$c^2$ which for a typical accretion 
efficiency $\eta=0.1$ equals $5 \times 10^{45}$ erg sec$^{-1}$ -- 
about a factor seven below the observed AGN luminosity of 3C\,68.2.  
An essential element in this discussion is the assumption that
the growth of the central black hole during AGN episodes is in phase
with the build-up of its host galaxy.  We cannot exclude the possibility
of a more energetic star-forming phase (for instance at SFR 
$\sim10^4$ M$_{\odot}$ yr$^{-1}$), but such SFRs remain to be observed 
in high-redshift radio galaxies.  However, if the star-forming phase 
lasts about a factor ten longer than the symbiotic phase, then 
$\sim$20 such events are needed to build-up the massive galaxy. 
In that case the mass consumption required to build its black hole 
during the $\sim$20 AGN episodes would be in agreement with the measured 
AGN luminosity.  Other forms of symbiotic occurrences can also be 
envisaged, including pure AGN phases not linked to coeval starbursts 
and the direct capture of stars through galaxy-galaxy merging.  
Whereas low redshift ULIRGs are believed to originate from major mergers, 
high redshift ULIRGs in the form of submm galaxies might have formed 
through cold gas accretion \citep{dave10};
the same has been postulated for high redshift AGN \citep{bourn11}. 
Even at their extreme stellar masses, our 3C hosts obey the
sSFR($z$) behaviour for non-AGN and X-ray AGN \citep{mullaney12a}.
However, their X-ray luminosities are higher, by orders of magnitudes;
moreover they are radio-loud. They may well mark the rare, most extreme
examples of the universal accretion to star-formation rate ratio
postulated by \citet{mullaney12b}. Following these authors, we contend
that the full symbiosis whereby a cold gas inflow driven, vigorous
star-formation episode leads into (and may be quenched by) an energetic
AGN phase, offers an attractive scenario for the reason of a common fuel
supply, building up both black hole and stellar mass. 

\section{Conclusions}

Following implications from submm emission, mid-IR spectra and Ly-$\alpha$
emission, we now know unambiguously that a substantial fraction of the
hosts of high-redshift radio galaxies are dust and gas rich, undergoing
one of likely many phases of copious star-formation.  Energetically
speaking they are comparable to ultra-luminous infrared galaxies.  Our
study shows accretion and star-formation activity simultaneously at work
in the most massive early-epoch galaxies, building these galaxies and
their central black holes, forcefully and effectfully. It will be
important to extend the investigations to radio-loud AGN hosted
by less massive galaxies. Radio-loud activity phases must play
an important role in galaxy formation.

\acknowledgments

The Herschel spacecraft was designed, built, tested, and launched under
a contract to ESA, managed by the Herschel/Planck Project team, by an
industrial consortium under the overall responsibility of the prime
contractor Thales Alenia Space (Cannes), and including Astrium
(Friedrichs-hafen) responsible for the payload module and for system
testing at spacecraft level, Thales Alenia Space (Turin) responsible for
the service module, and Astrium (Toulouse) responsible for the
telescope, with in excess of a hundred subcontractors.  HCSS/HSpot/HIPE
is a joint development by the Herschel Science Ground Segment
Consortium, consisting of ESA, the NASA Herschel Science Center, and the
HIFI, PACS and SPIRE consortia.  We acknowledge the support and interest
of the full Guaranteed Time Proposal team, of Dutch liaison astronomer
Max Avruch, of colleague Karina Caputi, and the comments of an
expert referee.

\begin{deluxetable}{lllllll}
\tabletypesize{\scriptsize}
\tablecaption{\scriptsize Observations and measured flux densities. Herschel observations
 using the PACS and SPIRE instruments yielded photometric data in five bands. The 1-sigma
errors combine the measurement noise and the background noise.}
\tablewidth{0pt}
\tablehead{
 \colhead{Object} & \colhead{RA(J2000)} & \colhead{DEC(J2000)} & \colhead{Obs.date} &
 \colhead{Obs.band} & \colhead{Integr.time} & \colhead{Flux density (mJy)}
 }
\startdata
3C\,368 & 18$^{\rm h}$05$^{\rm m}$06\fs36 & +11\arcdeg01\arcmin32\farcs5
                                             & 2011 March 22 & 70\micron  & 160sec & 30.0$\pm$2.0 \\
        &                   &                & 2011 March 22 & 160\micron & 160sec & 56.0$\pm$3.0 \\
        &                   &                & 2011 March 28 & 250\micron & 111sec & 31.1$\pm$7.0 \\
        &                   &                & 2011 March 28 & 350\micron & 111sec & 20.1$\pm$6.6 \\
        &                   &                & 2011 March 28 & 500\micron & 111sec & $<$21.0       \\
3C\,68.2 & 02$^{\rm h}$34$^{\rm m}$23\fs86 & +31\arcdeg34\arcmin17\farcs5
                                             & 2011 July 10  & 70\micron  & 160sec & 27.0$\pm$6.0 \\
        &                   &                & 2011 July 10  & 160\micron & 160sec & 40.0$\pm$7.0 \\
        &                   &                & 2011 July 31  & 250\micron & 111sec & 28.7$\pm$7.1 \\
        &                   &                & 2011 July 31  & 350\micron & 111sec & 29.6$\pm$7.4 \\
        &                   &                & 2011 July 31  & 500\micron & 111sec & 19.7$\pm$7.0 \\
3C\,257 & 11$^{\rm h}$23$^{\rm m}$09\fs17 & +05\arcdeg30\arcmin19\farcs5
                                             & 2011 June 1   & 70\micron  & 640sec & 11.6$\pm$1.0 \\
        &                   &                & 2011 June 1   & 160\micron & 640sec & 16.0$\pm$1.7 \\
        &                   &                & 2010 Nov 23   & 250\micron & 296sec & 22.8$\pm$4.9 \\
        &                   &                & 2010 Nov 23   & 350\micron & 296sec & 20.1$\pm$5.7 \\
        &                   &                & 2010 Nov 23   & 500\micron & 296sec & 20.4$\pm$6.8 \\
\enddata
\end{deluxetable}

\clearpage

\begin{deluxetable}{llllllllll}
\tabletypesize{\scriptsize}
\rotate
\tablecaption{\scriptsize AGN and star-formation luminosities. The inferred AGN and 
 star-formation luminosities for the three high-redshift radio galaxies are listed, as well
 as their (projected) radio source sizes and host galaxy masses, taken from the literature
 (with references). The 1-sigma errors were determined from the distribution of the ten
 best model combination sets.}
\tablewidth{0pt}
\tablecolumns{10}
\tablehead{
 \colhead{Object} & \colhead{Redshift} & \colhead{T$_{\rm cool}$} & \colhead{L$_{\rm AGN}$} &
 \colhead{L$_{\rm SF}$} & \colhead{L$_{\rm SF}$} & \colhead{SFR} & \colhead{Radio size} &
 \colhead{M$_{\rm host}$} & \colhead{M$_{\rm host}$/radio refs.}
 \\
 \colhead{} & \colhead{} & \colhead{(K)} & \colhead{(erg sec$^{-1}$)} &
 \colhead{(erg sec$^{-1}$)} & \colhead{($10^{12}$ L$_{\sun}$)} &
 \colhead{(M$_{\sun}$ yr$^{-1}$)} & \colhead{(kpc)} & \colhead{(M$_{\sun}$)} & \colhead{}
 }
\startdata
3C\,368  & 1.132 & 53$\pm$1  & (7.3$\pm$3.5)$\times10^{45}$ & (1.4$\pm$0.04)$\times10^{46}$
  & 3.5$\pm$0.1 & 610 & 80  & $10^{11.6}$ & \citet{best98a, best98b} \\
3C\,68.2 & 1.575 & 36$\pm$3 & (3.3$\pm$1.0)$\times10^{46}$ & (8.6$\pm$0.8)$\times10^{45}$
  & 2.2$\pm$0.2 & 390 & 190 & $10^{11.7}$ & \citet{best97, best98a} \\
3C\,257  & 2.474 & 37$\pm$3 & (4.1$\pm$2.0)$\times10^{46}$ & (1.7$\pm$0.1)$\times10^{46}$
  & 4.4$\pm$0.3 & 770 & 95  & $<10^{11.7}$ & \citet{breuck10, breugel98} \\
\enddata
\end{deluxetable}

\begin{figure}
 \epsscale{0.60} 
 \plotone{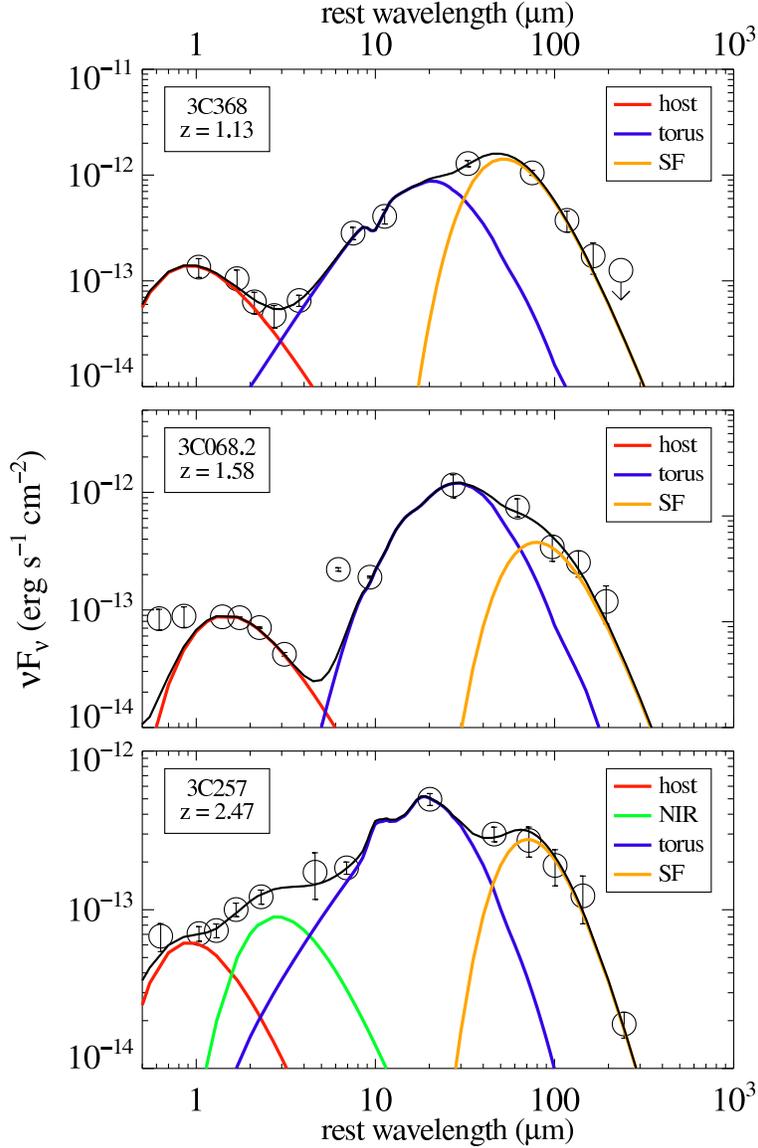}
 \caption{Infrared-submm spectral energy distributions.
  The infrared-submm SEDs of 3C\,368, 3C\,68.2, and 3C\,257 are shown,
  with the best-fitting, multi-component model superposed: stellar
  radiation (red), torus radiation (blue), and cool dust radiation
  (yellow).  3C\,257 requires an extra hot dust component (green) -- see
  the main text.  The mismatch at 16$\micron$ ~for 3C\,68.2 is possibly due
  to luminous PAH-emission. The error bars reflect $\pm 1\sigma$ errors.}
\end{figure}

\end{document}